\documentclass[onecolumn,preprintnumbers,pra,amsfonts,amsmath,showkeys,nofootinbib,10pt]{revtex4}
\usepackage{amsfonts}
\usepackage{amsmath}
\usepackage{graphicx}
\usepackage{epsfig}		
\usepackage{bm}
\newcommand{\be}{\begin{equation}}
\newcommand{\ee}{\end{equation}}
\newcommand{\br}{\begin{eqnarray}}
\newcommand{\er}{\end{eqnarray}}

\newcommand{\bd}{\begin{displaymath}}
\newcommand{\ed}{\end{displaymath}}

\newcommand{\bfig}{\begin{figure}}
\newcommand{\efig}{\end{figure}}
\newcommand{\SE}{Schr\"{o}dinger equation}

\begin{document}

\title{Many-particle Sudarshan-Lindblad equation: mean-field approximation, nonlinearity and dissipation in a spin system}
\author{G. A. Prataviera$^{1}$ and S. S. Mizrahi$^{2}$}
\email{prataviera@usp.br, salomon@df.ufscar.br}
\affiliation{$^{1}$Departamento de Administra\c{c}{\~a}o, FEA-RP, Universidade de S{\~a}%
o Paulo, 14040-905, Ribeir\~{a}o Preto, SP, Brazil \\
$^{2}$Departamento de F\'{\i}sica, CCET, Universidade Federal de S\~ao Carlos, 
13565-905, S\~ao Carlos, SP, Brazil}

\begin{abstract}
A system of $N$ spin-1/2 particles interacting with a thermal reservoir is
used as a pedagogical example for advanced undergraduate and graduate
students. We introduce and illustrate some methods, approximations, and
phenomena related to dissipation and nonlinearity in many-particle physics. We
start our analysis from the dynamical Sudarshan-Lindblad quantum master
equation for the density operator of a system $\mathcal{S}$ interacting with a
thermal reservoir $\mathcal{R}$. We derive the quantum version of the
so-called Bogoliubov-Born-Green-Kirkwood-Yvon (BBGKY) equations such that the
master equation can be decomposed in a hierarchical set of $N-1$ equations
($N>1$). The hierarchy is broken by introducing the mean-field approximation
and reducing the problem to a nonlinear single particle system. In this
scenario, the Hamiltonian is nonlinear (\emph{i.e.}, it depends on the state of
$\mathcal{S}$), although the superoperator responsible for the dissipation and
decoherence of $\mathcal{S}$ remains unaffected. To provide a useful tool to
students: (1) we discuss the physical approximations involved, (2) we derive
the analytical solution to the mean values equations of motion resulting from
the Hamiltonian, (3) we solve analytically the master equation in the
stationary regime, (4) we obtain and discuss the solution of the nonlinear
master equation, numerically, and finally, (5) we discuss the master equation
beyond the mean-field approximation and show how to introduce higher order
quantum correlations that have been previously neglected.

\end{abstract}

\maketitle




%
\section{Introduction}
%
Since the birth of quantum mechanics (QM) the investigation of irreversible processes 
has received special emphasis, with the aim of describing a physical system 
evolving in compliance with the human feeling that there exists an ``arrow of time'' 
in nature. The approaches had to go along with the observed physical systems where 
macroscopic phenomena evolve inexorably to an equilibrium or stationary state, so breaking the time symmetric 
microreversibility present in the dynamical \SE \footnote{That means that by changing $t$ by $-t$, reverting the 
velocities, and replacing the initial conditions in the solution of the 
equation by the final ones does not lead us to the initial conditions after 
a time $t$ has elapsed.}. 
Since its incipience several thinkers have devoted efforts to 
describe the irreversible processes with the care to avoid the violation of the QM principles. 
For a few historical reading see the papers \cite{Land27,Bloch28,Pauli28,WW30,Bate31,Cal,kanai,
vanHove57,havas,Zwan58,Toda58,kerner}, and for a review we recommend the books 
\cite{haake73,louisell73,Dav76,Breu02}. 

The standard rationale to describe formally the time irreversibility of a system of interest  $\mathcal{S}$ 
is to couple it to another one, $\mathcal{R}$, 
chosen according to the phenomenon one wants to observe -- such a system 
is sometimes called \emph{the rest of the universe},  
\emph{the environment} or \emph{the reservoir} -- and make them to interact 
via a convenient potential. Thence, as times go on, both systems will 
evolve and eventually intertwine their degrees of freedom and will 
share a common quantum state. However, since one does not have access 
(either by observation or through experiment) to the details of 
$\mathcal{R}$, methodologically one proceeds as one commonly does in 
statistical theory: one gets the reduced probability function for 
$\mathcal{S}$ by integrating (or summing) over the variables representing 
the rest of the universe $\mathcal{R}$. In quantum mechanical language one 
says that we calculate the trace over the system $\mathcal{R}$ degrees of freedom.
So one remains with a dynamical equation, called master equation, for the 
state of the system $\mathcal{S}$, represented by the density operator
(or density matrix), which shall evolve irreversibly in time as originally sought. The master equation for 
$\mathcal{S}$ will contain the influence of $\mathcal{R}$ through 
the presence of extra terms depending on its parameters. This approach 
received a thorough attention essentially in the 1960's, as can be seen in the researches reported in Refs. \cite{sen60,schwin61,zwanzig64,fordkac65,louwal65,opp65,weid65,lax65,uller66,bonif67,louimar67,lambscul67,Belav,peier72,Kraus71,agar73}, and for some more recent applications see Refs. \cite{carm,gar,milb,salo,gamble,mizel}. In addition, in classical statistical mechanics \cite{reif} and in the stochastic processes literature \cite{vankampen}, a master equation is a gain-loss equation, in terms of probability transitions, for the probabilities of the possible states of a system, while in quantum mechanics, the master equation governs the density operator evolution in time \cite{carm}. 

The standard procedure to describe irreversibility in $\mathcal{S}$, starting 
from equations that are time-reversible for the complete system $\mathcal{S}+\mathcal{R}$,
consists in deducing a differential equation for the $\mathcal{S}$ state only,
by calculating the trace over the degrees of freedom or washing out the detailed information
about $\mathcal{R}$, which leaves us with a new new modified (time-irreversible) equation 
for $\mathcal{S}$ containing the parameters that were already present in $\mathcal{R}$ and also news one as the 
absolute temperature. This approach leads to an essential 
equation known as Sudarshan-Lindblad (SL) equation \cite{Kossa,gori76,lind76}, that describes irreversibility, 
decoherence, and dissipation in $\mathcal{S}$. In addition, this approach has been applied 
to several recent papers related to the quantum measurement theory \cite{percival,peres,cresser,brasil,yan}. 
A recent paper was devoted to the derivation of that equation with the pedagogical purpose of being accessible for undergraduate students in physics, chemistry and mathematics, having a basic background on QM \cite{BFN}. 

In this paper we have the concern to be quite pedagogical in order to assist advanced undergraduate and graduate students interested in many-particle physics, once it involves irreversibility, dissipation and nonlinearity for a many-particle system. We show that for a system  $\mathcal{S}$ constituted by N free spin-$1/2$ particles (or, by the mathematical 
isomorphism, a system of two-level atoms) interacting with a thermal reservoir $\mathcal{R}$, and adopting the mean-field approximation, the single-particle Sudarshan-Lindblad equation remains the same, although the Hamiltonian acquires, additionally, nonlinear terms, i.e., they depend on the state of the system $\mathcal{S}$, and the dynamical equations for the spin operators mean values become nonlinear. We obtain and digress on the solution of the nonlinear equations of the single-particle mean values operators, and how to go beyond the mean-field approximation by introducing two-particle quantum correlations.

The article is organized as follows: In Sec. II we introduce the Hamiltonian and the master equation describing
the system of N spin-1/2 particles interacting with a thermal reservoir. In Sec. III, a mean-field
approximation is developed and an effective single-particle Hamiltonian is
obtained. In Sec. IV, we derive the analytical solution for the nonlinear mean-values operators, whose
motion is generated by the nonlinear Hamiltonian without dissipative terms, and illustrate
numerically the case with dissipation.  In Sec V, we discuss how to go beyond the
mean-field approximation by introducing two-particle correlations. Finally, in Sec. VI we present a summary and the
conclusions.

%
\section{Evolution equation for a $N$ spin-1/2 particles interacting with a reservoir}
We consider dissipation in physical systems in the framework of the System-Reservoir interaction. In this framework, the full Hamiltonian describing a generic system of interest
$\mathcal{S}$ interacting with a reservoir $\mathcal{R}$ in thermal equilibrium, is given by
\begin{eqnarray}
H_{\mathcal
{S}+\mathcal{R}}=H_{\mathcal{S}}+
H_{\mathcal{R}}+H_{\mathcal{SR}},
\label{hamil}
\end{eqnarray}
where $H_{\mathcal{S}}$ is the system of interest Hamiltonian, $
H_{\mathcal{R}}$ is the reservoir Hamiltonian, and $H_{\mathcal{SR}}$ is the system-reservoir interaction Hamiltonian. As shown in Refs \cite{BFN,carm,gar,milb}, by tracing over the reservoir degrees of freedom, considering a bilinear interaction between system and reservoir, and under the Markov approximation, the system density operator $\rho_{\mathcal{S}}$ has its evolution described by the Sudarshan-Lindblad master equation
\begin{equation}
\label{suda}\frac{\partial\rho_{\mathcal{S}}}{\partial t}
=-\, i[H_{\mathcal{S}},\rho_{\mathcal{S}}]
-\frac{(\bar{n}+1)\Gamma}{2}\mathcal{L}[\hat{A}]\rho_{\mathcal{S}}
- \frac{\bar{n}\Gamma}{2}\mathcal{L}[\hat{A}^{\dagger}]\rho_{\mathcal{S}}.
\end{equation}
The superoperator $\mathcal{L}[\hat{A}]$ is defined for an arbitrary system operator $\hat{A}$ by
\begin{equation}
\mathcal{L}[\hat{A}]\rho_{\mathcal{S}}=[\hat{A},\hat{A}^{\dagger}\rho_{\mathcal{S}}]-[\hat{A}^{\dagger},\rho_{\mathcal{S}}\hat{A}],
\end{equation}
where $\Gamma $ is the decay rate constant, and $\bar{n}$ is the mean number of the
reservoir quanta. In the right hand side (RHS) of Eq. (\ref{suda}), the first term describes the unitary evolution of $\mathcal{S}$, while the last two terms are responsible for the non-unitary irreversible and dissipative evolution of $\mathcal{S}$. By particular choices of $H_{\mathcal{S}}$ and the operator $\hat{A}$, the Eq. (\ref{suda}) has been applied to several problems related to dissipation in quantum optics and in measurement theory \cite{carm,gar,milb}.

Here, we are interested in describing a system consisting of $N$ spin-1/2 particles (or two-level atoms, since their description accepts the same algebra), interacting with a
reservoir, and whose constituents do not interact with each other or, because their density is quite low their
direct interaction can be neglected. In our case, the Hamiltonian (\ref{hamil}) describing $N$ particles interacting with the reservoir (an infinite number of harmonic oscillators) is given by \cite{louisell73}
\begin{eqnarray}
 H_{\mathcal{S+R}}=\frac{\omega_{o}}{2} S_{0} + \sum_{k}\omega_{k} b_{k}^{\dagger}b_{k}+  
 \sum_{k}(g_{k}b_{k}S_{+}+g_{k}^{*} b_{k}^{\dagger}S_{-}),
\label{primeira}
\end{eqnarray}
where
\begin{equation}\label{opdick}
S_{\pm}=\sum_{i=1}^{N}\sigma_{\pm}(i), \hspace{1cm} S_{0}=\sum_{i=1}^{N}\sigma_{0}(i),
\end{equation}
%
%
are the collective operators which satisfy the following commutation relations
%
\begin{eqnarray}
&& [S_{0},S_{\pm}]=\sum_{i,i'=1}^{N}\left[ \sigma_{0}(i),\sigma_{\pm}(i')\right] = 2\sum_{i,i'=1}^{N} \sigma_{\pm}(i)\delta_{ii'}=\pm 2S_{+}, \label{cdk1}\\
&&[S_{+},S_{-}]=\sum_{i,i'=1}^{N}[{
\sigma}_{+}(i),\sigma_{-}(i')]= \sum_{i,i'=1}^{N} \sigma_{0}(i)\delta_{ii'}=S_{0},\label{cdk3}
\end{eqnarray}
%
and, the single-particle spin operators for the particle $i$ are $\sigma_{0}(i)$ and $\sigma_{\pm}(i)$;\footnote{With $\sigma_{\pm}= \frac{1}{2}\left(\sigma_{x} \pm i \sigma_{y}\right)$, where $\sigma_{x}$, $\sigma_{y}$, and $\sigma_{0}$ are represented by the Pauli matrices.}  $\omega_{0}$ is the transition frequency between the particles' two-states. The operators $b^{\dagger}_{k}$ and $b_{k}$ obey the commutation relation $[b_{k},b^{\dagger}_{k'}]=\delta_{k k'}$; their r\^{o}le is to create and annihilate a quantum of energy, $\hbar \omega_{k}$, of the reservoir $\mathcal{R}$; $\omega_{k}$ is the angular frequency associated to $k$-th harmonic oscillator belonging to $\mathcal{R}$. For the system-reservoir interaction, it is assumed that each oscillator interacts with all the particles with the same strength $ |g_ {k}| $. So, the master equation structure for the set of particles is identical to the one corresponding to the single-particle master equation \cite{louisell73} where the operators $\sigma_ {\pm}, \sigma_ {0} $ are replaced by the collective ones $S_ {\pm}, S_ {0} $. Then, identifying the operator $\hat{A}$, in Eq. (\ref{suda}), with the collective operator $S_ {+}$, the Sudarshan-Lindblad $N$ particles master equation acquires the form
\begin{eqnarray}
\label{EMDickT}\frac{\partial\rho_{N}}{\partial t}
=-\, i\frac{\omega_{0}}{2}[S_{0},\rho_{N}]
-\frac{(\bar{n}+1)\Gamma}{2V}\left([S_{+},S_{-}{\rho}_{N}]- \right. 
\left. [S_{-},\rho_{N}
S_{+}]\right)
- \frac{\bar{n}\Gamma}{2V}\left([S_{-},{
S}_{+}\rho_{N}]-[S_{+},\rho_{N}{
S}_{-}]\right),
\end{eqnarray}
where
\begin{eqnarray}\label{rhoN}
{\rho}_{N}\equiv{\rho}(1,2,3,...,N-1,N),
\end{eqnarray}
is the $N$-particle density operator, $V$ is the volume of the container, $\bar{n}=1/\left(e^{\hbar\omega_{o}/k_{B}T}-1 \right)$ is the mean number of the reservoir quanta at temperature $T$, and $k_{B}$ is the Boltzmann constant.

\section{The nonlinear master equation}{\label{section3}}
%
From Eq. (\ref{EMDickT}) we will deduce a quantum equivalent to the Bogoliubov-Born-Green-Kirkwood-Yvon (BBGKY) hierarchy \cite{liboff,mcquarrie} of equations similar to the procedure in classical kinetic theory. The term hierarchy refers to the order of correlations between $K$ particles, assumed as an approximation to the exact $N$-particle master equation ($N>K$). For example, an equation of evolution for the state of $K$ particles will depend on the state of $K+1$ particles, with $K+1\leq N$. The hierarchy is said to be broken whenever the state of $K+1$ particles is reduced as a sum of product of states involving less particles, \emph{i.e.}, introducing lower-order correlations.

Let us consider a subsystem constituted of $K$ particles ($K<N$), whose density operator is obtained by calculating
the trace over the remaining $K+1,K+2,...,N$ particles in Eq. (\ref{rhoN}). Then, the reduced density operator is
written as
\begin{equation} \label{rhoK}
\rho_{K}=Tr_{K+1,...,N}\left(\rho_{N}\right).
\end{equation}
The equation of motion for $\rho _{K}$ is obtained by calculating trace over
the master equation (\ref{EMDickT}), thus obtaining explicitly,


\begin{eqnarray}\label{masterKterCla}
\frac{\partial{\rho_{K}}}{\partial
t} &=& -i\, Tr_{K+1,...,N}\left\{\frac{\omega_{o}}{2}[S_{0},\rho_{N}]\right\}
-\frac{\Gamma}{2V}(\bar{n}+1)\left\{Tr_{K+1,...,N}\left([S_{+},S_{-}\rho_{N}]\right)
+Tr_{K+1,...,N}\left([\rho_{N}S_{+},S_{-}]\right)\right\} \nonumber\\
&&-\frac{\Gamma}{2V}\bar{n}\left\{Tr_{K+1,...,N}\left([S_{-},
S_{+}\rho_{N}]\right)+Tr_{K+1,...,N}\left([\rho_{N}S_{-},S_{+}]\right)\right\}.
\end{eqnarray}
The RHS of Eq.\eqref{masterKterCla} contains terms having the following structures
%
\begin{eqnarray}
&Tr_{K+1,...,N}\left([S_{\lambda},\rho_{N}]\right)\label{traco1}\\
&Tr_{K+1,...,N}\left([S_{\lambda},S_{\lambda'}
\rho_{N}]\right)\label{traco2}\\
&Tr_{K+1,...,N}\left([S_{\lambda},\rho_{N}{
S}_{\lambda'}]\right),\label{traco3}
\end{eqnarray}
%
with $\lambda , \lambda' = -,0,+$. Detailing the calculations, the term (\ref{traco1}) can be written as
\begin{eqnarray}\label{tra1}
Tr_{K+1,...,N}\left([S_{\lambda},\rho_{N}]\right)&=&
Tr_{K+1,...,N}\left(\sum_{i=1}^{N}[{
\sigma}_{\lambda}(i),\rho_{N}]\right)\nonumber\\
&=&\sum_{i=1}^{K}[\sigma_{\lambda}(i),Tr_{K+1,...,N}\left(\rho_{N}\right)]
+\sum_{i=K+1}^{N}Tr_{K+1,...,N}\left([{
\sigma}_{\lambda}(i),\rho_{N}]\right)\nonumber\\
&=&\sum_{i=1}^{K}[\sigma_{\lambda}(i),\rho_{K}]+\sum_{i=K+1}^{N}Tr_{K+1,...,N}\left([{
\sigma}_{\lambda}(i),\rho_{N}]\right),
\end{eqnarray}
where the sum $\sum_{i=1}^{N}...$ is broken into two terms, $\sum_{i=1}^{K}...+\sum_{i=K+1}^{N}...$, since the trace operation is done over the particles $K+1,...,N$.
The second term in the RHS of the last line in Eq. \eqref{tra1} is
\begin{eqnarray}\label{tra1a}
\sum_{i=K+1}^{N}Tr_{K+1,...,N}\left([{
\sigma}_{\lambda}(i),\rho_{N}]\right)&=Tr_{K+1,...,N}\left([{
\sigma}_{\lambda}(K+1)+...+{
\sigma}_{\lambda}(N),\rho_{N}]\right),
\end{eqnarray}
where all the $N-K$ terms are equivalent, since the particles are assumed being identical. Therefore, using the cyclic property
of the trace operation ($Tr({ A}{ B})=Tr({ B}{ A})$), it follows that
\begin{eqnarray}\label{tra1b}
\sum_{i=K+1}^{N}Tr_{K+1,...,N}\left([\sigma_{\lambda}(i),\rho_{N}]\right)
=(N-K)Tr_{K+1,...,N}\left([\sigma_{\lambda}(K+1),\rho_{K+1}]\right) = 0.
\end{eqnarray}
Then, Eq. \eqref{tra1} reduces to
\begin{equation}\label{tra1c}
Tr_{K+1,...,N}\left([{S}_{\lambda},\rho_{N}]\right)
=\sum_{i=1}^{K}[\sigma_{\lambda}(i),\rho_{K}].
\end{equation}

Now, considering the trace operation \eqref{traco2}, and breaking again the sum $\sum_{i=1}^{N}...$ as $\sum_{i=1}^{K}...+\sum_{i=K+1}^{N}...$, we obtain the following four terms
\begin{eqnarray}\label{tra2}
Tr_{K+1,...,N}\left([S_{\lambda},{
S}_{\lambda'}\rho_{N}]\right)\!&=&\!
Tr_{K+1,...,N}\left(\sum_{i,i'=1}^{N}
[\sigma_{\lambda}(i),\sigma_{\lambda'}(i')\rho_{N}]\right)\nonumber\\
\!&=&\! Tr_{K+1,...,N}\left(\sum_{i,i'=1}^{K}
[\sigma_{\lambda}(i),\sigma_{\lambda'}(i')\rho_{N}]\right)\nonumber\\
\!&+&\!Tr_{K+1,...,N}\left(\sum_{i=1}^{K}\sum_{i'=K+1}^{N}
[\sigma_{\lambda}(i),\sigma_{\lambda'}(i')\rho_{N}]\right)\nonumber\\
\!&+&\! Tr_{K+1,...,N}\left(\sum_{i=K+1}^{N}\sum_{i'=1}^{K}
[\sigma_{\lambda}(i),\sigma_{\lambda'}(i')\rho_{N}]\right)\nonumber\\
\!&+&\!Tr_{K+1,...,N}\left(\sum_{i,i'=K+1}^{N}
[\sigma_{\lambda}(i),\sigma_{\lambda'}(i')\rho_{N}]\right)\, .
\end{eqnarray}
Evaluating explicitly each one separately we get:

\begin{eqnarray}\label{resultra2a}
\textbf{(1)}\ Tr_{K+1,...,N}\left(\sum_{i,i'=1}^{K}
[\sigma_{\lambda}(i),\sigma_{\lambda'}(i')\rho_{N}]\right)=\sum_{i,i'=1}^{K}
[\sigma_{\lambda}(i),\sigma_{\lambda'}(i')Tr_{K+1,...,N}\left(\rho_{N}\right)]
=\sum_{i,i'=1}^{K}
[\sigma_{\lambda}(i),\sigma_{\lambda'}(i')\rho_{K}];\nonumber\\
&&
\end{eqnarray}

\begin{eqnarray}\label{resultra2b}
\textbf{(2)}\ Tr_{K+1,...,N}\left(\sum_{i=1}^{K}\sum_{i'=K+1}^{N}
[\sigma_{\lambda}(i),\sigma_{\lambda'}(i')\rho_{N}]\right)&=&\sum_{i=1}^{K}\sum_{i'=K+1}^{N}
[\sigma_{\lambda}(i),Tr_{K+1,...,N}\left(\sigma_{\lambda'}(i')\rho_{N}\right)]\nonumber\\
&=&(N-K)\sum_{i=1}^{K}[\sigma_{\lambda}(i),Tr_{K+1}\left(\sigma_{\lambda'}(K+1)\rho_{K+1}\right)],\nonumber\\
&&
\end{eqnarray}
where the factor $N-K$ is present because there are formally equivalent terms;

\begin{eqnarray}\label{resultra2c}
\textbf{(3)}\ Tr_{K+1,...,N}\left(\sum_{i=K+1}^{N}\sum_{i'=1}^{K}
[\sigma_{\lambda}(i),\sigma_{\lambda'}(i')\rho_{N}]\right)&=\sum_{i'=1}^{K}
\sigma_{\lambda'}(i')\sum_{i=K+1}^{N}Tr_{K+1,...,N}
\left([\sigma_{\lambda}(i),\rho_{N}]\right)=0;\nonumber\\
&&
\end{eqnarray}

\begin{eqnarray}\label{resultra2d}
\textbf{(4)}\ Tr_{K+1,...,N}\left(\sum_{i,i'=K+1}^{N}
[\sigma_{\lambda}(i),\sigma_{\lambda'}(i')\rho_{N}]\right)
&=\sum_{i,i'=K+1}^{N}Tr_{K+1,...,N}
\left([\sigma_{\lambda}(i),\sigma_{\lambda'}(i')\rho_{N}]\right)=0.\nonumber\\
&&
\end{eqnarray}
In Eqs. (\ref{resultra2c}) and (\ref{resultra2d}) the results are zero because the trace of a commutator is null,
then Eq. \eqref{tra2} becomes
\begin{eqnarray}\label{traco2sol}
Tr_{K+1,...,N}\left([S_{\lambda},{
S}_{\lambda'}\rho_{N}]\right)=\sum_{i,i'=1}^{K}
[\sigma_{\lambda}(i),\sigma_{\lambda'}(i')\rho_{K}]
+(N-K)\sum_{i=1}^{K}[\sigma_{\lambda}(i),Tr_{K+1}
\left(\sigma_{\lambda'}(K+1)\rho_{K+1}\right)].\nonumber\\
&&
\end{eqnarray}
Analogously to the above procedure, for the Eq. \eqref{traco3} we have
\begin{eqnarray}\label{tra3}
Tr_{K+1,...,N}\left([S_{\lambda},\rho_{N}{
S}_{\lambda'}]\right)&=& Tr_{K+1,...,N}\left(\sum_{i,i'=1}^{N}
[\sigma_{\lambda}(i),\rho_{N}\sigma_{\lambda'}(i')]\right)\nonumber\\
&=&Tr_{K+1,...,N}\left(\sum_{i,i'=1}^{K}
[\sigma_{\lambda}(i),\rho_{N}\sigma_{\lambda'}(i')]\right)\nonumber\\
&+&Tr_{K+1,...,N}\left(\sum_{i=1}^{K}\sum_{i'=K+1}^{N}
[\sigma_{\lambda}(i),\rho_{N}\sigma_{\lambda'}(i')]\right)\nonumber\\
&+&Tr_{K+1,...,N}\left(\sum_{i=K+1}^{N}\sum_{i'=1}^{K}
[\sigma_{\lambda}(i),\rho_{N}\sigma_{\lambda'}(i')]\right)\nonumber\\
&+&Tr_{K+1,...,N}\left(\sum_{i,i'=K+1}^{N}
[\sigma_{\lambda}(i),\rho_{N}\sigma_{\lambda'}(i')]\right)\, ,\nonumber\\
\end{eqnarray}
which results again in four terms. Looking at each one separately and following the same procedure as above, we get

\begin{eqnarray}\label{resultra3a}
\textbf{(1')}\ Tr_{K+1,...,N}\left(\sum_{i,i'=1}^{K}
[\sigma_{\lambda}(i),\rho_{N}\sigma_{\lambda'}(i')]\right)=\sum_{i,i'=1}^{K}
[\sigma_{\lambda}(i),Tr_{K+1,...,N}\left(\rho_{N}\right)\sigma_{\lambda'}(i')]
=\sum_{i,i'=1}^{K}
[\sigma_{\lambda}(i),\rho_{K}\sigma_{\lambda'}(i')], \nonumber \\
&&
\end{eqnarray}

\begin{eqnarray}\label{resultra3b}
\textbf{(2')}\ Tr_{K+1,...,N}\left(\sum_{i=1}^{K}\sum_{i'=K+1}^{N}
[\sigma_{\lambda}(i),\rho_{N}\sigma_{\lambda'}(i')]\right)&=&\sum_{i=1}^{K}\sum_{i'=K+1}^{N}
[\sigma_{\lambda}(i),Tr_{K+1,...,N}\left(\rho_{N}\sigma_{\lambda'}(i')\right)]\nonumber\\
&=&(N-K)\sum_{i=1}^{K}[\sigma_{\lambda}(i),Tr_{K+1}\left(\rho_{K+1}\sigma_{\lambda'}(K+1)\right)], \nonumber \\
&&
\end{eqnarray}

\begin{eqnarray}\label{resultra3c}
\textbf{(3')}\ Tr_{K+1,...,N}\left(\sum_{i=K+1}^{N}\sum_{i'=1}^{K}
[\sigma_{\lambda}(i),\rho_{N}\sigma_{\lambda'}(i')]\right)&=\sum_{i=K+1}^{N}Tr_{K+1,...,N}
\left([\sigma_{\lambda}(i),\rho_{N}]\right)\sum_{i'=1}^{K}
\sigma_{\lambda'}(i')=0, \nonumber \\
&&
\end{eqnarray}

\begin{eqnarray}\label{resultra3d}
\textbf{(4')}\ Tr_{K+1,...,N}\left(\sum_{i,i'=K+1}^{N}
[\sigma_{\lambda}(i),\rho_{N}\sigma_{\lambda'}(i')]\right)
&=\sum_{i,i'=K+1}^{N}Tr_{K+1,...,N}
\left([\sigma_{\lambda}(i),\rho_{N}\sigma_{\lambda'}(i')]\right)=0. \nonumber \\
&&
\end{eqnarray}
Therefore, Eq. \eqref{tra3} can be written as
\begin{eqnarray}\label{traco3sol}
Tr_{K+1,...,N}\left([S_{\lambda},\rho_{N}{
S}_{\lambda'}]\right)=\sum_{i,i'=1}^{K}
[\sigma_{\lambda}(i),\rho_{K}\sigma_{\lambda'}(i')]
+(N-K)\sum_{i=1}^{K}[\sigma_{\lambda}(i),Tr_{K+1}
\left(\rho_{K+1}\sigma_{\lambda'}(K+1)\right)].\nonumber\\
&&
\end{eqnarray}
Thus, back to Eqs. (\ref{traco1})-(\ref{traco3}), we obtain explicitly
%
\begin{eqnarray}
Tr_{K+1,...,N}\left([S_{\lambda},\rho_{N}]\right)
&=&\sum_{i=1}^{K}[\sigma_{\lambda}(i),\rho_{K}]\label{tracofinal1},\\
Tr_{K+1,...,N}\left([S_{\lambda},S_{\lambda'}
\rho_{N}]\right)&=&\sum_{i,i'=1}^{K}
[\sigma_{\lambda}(i),\sigma_{\lambda'}(i')\rho_{K}]+(N-K)\sum_{i=1}^{K}[\sigma_{\lambda}(i),Tr_{K+1}
\left(\sigma_{\lambda'}(K+1)\rho_{K+1}\right)]\label{tracofinal2}, \nonumber \\
\\
Tr_{K+1,...,N}\left([S_{\lambda},\rho_{N}{
S}_{\lambda'}]\right)&=&\sum_{i,i'=1}^{K}
[\sigma_{\lambda}(i),\rho_{K}\sigma_{\lambda'}(i')]+(N-K)\sum_{i=1}^{K}[\sigma_{\lambda}(i),Tr_{K+1}
\left(\rho_{K+1}\sigma_{\lambda'}(K+1)\right)]. \nonumber \\
\label{tracofinal3}
\end{eqnarray}
%

Using the results in Eqs. (\ref{tracofinal1})-(\ref{tracofinal3}) we obtain the following master equation for the reduced $K$-particle system
\begin{eqnarray}\label{masterKterClaF}
\frac{\partial{\rho_{K}}}{\partial
t}&=&-i\frac{\omega_{o}}{2}\sum_{i=1}^{K}[\sigma_{0}(i),\rho_{K}] -
(N-K)\frac{\Gamma}{2V}\sum_{i=1}^{K}\left(
[\sigma_{+}(i),Tr_{K+1}
\left(\sigma_{-}(K+1)\rho_{K+1}\right)]+h.c.
\right) \nonumber\\
&& -\frac{\Gamma}{2V}\sum_{i,i'=1}^{K}\left((\bar{n}+1)
[\sigma_{+}(i),\sigma_{-}(i')\rho_{K}]
+\bar{n}
[\sigma_{-}(i),\sigma_{+}(i')\rho_{K}]+h.c.\right),
\end{eqnarray}
with $K=1,2,...,N-1$, and where $h.c.$ stands for hermitian conjugate. It is worth noting that the master equation for the reduced density operator of $K$ particles $\rho_{K}$ depends on the density operator of $K+1$ particles, $\rho_{K+1}$. These $N-1$ equations are similar to the BBGKY hierarchy, studied in several contexts related to the statistical mechanics of classical fluids \cite{liboff,mcquarrie}. For a single representative particle of the system ($K=1$), the equation of motion for $\rho _{1}$ is

\begin{eqnarray}\label{masterK1}
\frac{\partial{\rho_{1}}}{\partial t}&=&
-i\frac{\omega_{o}}{2}[\sigma_{0},\rho_{1}]\!-\!\frac{(N-1)\Gamma}{2V}\left(
[\sigma_{+},Tr_{2}\sigma_{-}(2)\rho_{2}] +h.c.\right)\nonumber\\
 && - \frac{(\bar{n}+1)\Gamma}{2V}\left([\sigma_{+},\sigma_{-}\rho_{1}] +h.c.\right)
\!- \!\frac{\bar{n}\Gamma}{2V}\left( [\sigma_{-},\sigma_{+}\rho_{1}] +h.c.\right),
\end{eqnarray}


\noindent noting that it depends on the two-particle density operator $\rho_{2}$, and so on for the whole hierarchy.
Thus, we still have a complex problem. In order to obtain a scenario easier to be
handled mathematically, some kind of approximation needs to be done. A plausible physical
argument to obtain some simplification for a dilute system consists in disregarding the higher order
particle correlations. This is achieved when one does the factorization
$\rho_{2}(1,2)=\rho_{1}(1)\rho_{1}(2)$, so now a generic particle is assumed to move in a mean-field
produced by all the other particles. This approximation is very similar to the well-known Hartree
approximation or mean-field approximation used in atomic and condensed matter physics \cite{gasi,eis,antunes}, which consist in a factorization as a product of single-particle states.
Implementing this approximation, Eq. (\ref
{masterK1}) reduces to a generic single-particle master equation

\begin{eqnarray}
\label{nlTerCla}
\frac{\partial\rho_{1}}{\partial t}=-i[H_{ef},\rho_{1}]-
\frac{\Gamma}{2V}\left((\bar{n}+1)[
\sigma_{+},\sigma_{-}\rho_{1}]+\right. 
\left.  \bar{n}[
\sigma_{-},\sigma_{+}\rho_{1}]+h.c.\right),
\end{eqnarray}
where a single-particle effective hermitian Hamiltonian is identified as
\begin{equation}\label{Hef2}
{ H}_{ef}=\frac{\omega_{o}}{2}\sigma_{0} +i\frac{(N-1)\Gamma}{2V}\left[\langle\sigma_{+}\rangle\sigma_{-} -\langle\sigma_{-}\rangle\sigma_{+}\right],
\end{equation}
which contains nonlinear terms, obtained by grouping the second term in the RHS of Eq. (\ref{masterK1}) to the original commutator, namely $Tr_{2}\left(\sigma_{\pm}(2)\rho_{2}\right)=\rho_1(1) Tr_{2} \left(\sigma_{\pm}(2)\rho_{1}(2)
\right )= \rho_1(1)\left\langle \sigma_{\pm}\right\rangle$, reminding that $Tr(\rho \hat{A})\equiv \langle \hat{A}\rangle$ is the mean value of an operator $\hat{A}$ in the density operator formalism. The second term in the brackets in the Hamiltonian (\ref{Hef2}) represents a particle acted by an effective polarization field $E_{pol}=i(N-1)\Gamma\langle\sigma_{+}\rangle/2V$ originated from the other $(N-1)$ particles that produce the mean-field effect, proportional to the uniform particle density $(N-1)/V$. In this approximation, the structure of the
Sudarshan-Lindblad dissipative terms remains the same, although the
single-particle Hamiltonian acquires, additional nonlinear terms.

\section{Dynamics}{\label{section4}}

By noting that
\begin{equation}
\frac{d}{dt}\langle \sigma _{i}\rangle=\frac{d}{dt} Tr  \left(\sigma _{i}\rho_{1}\right) =Tr( \sigma _{i} \dot{\rho_{1}}),\ \ i= 0,\pm.
\end{equation}
the equations of motion for the single-particle operators mean values are obtained by inserting the RHS of Eq. (\ref{nlTerCla}) for $\dot{\rho_{1}}$. Then, we obtain the following nonlinear equations of motion for the volumetric density operators mean values,
\begin{eqnarray}\label{s0}
\frac{d}{dt}\langle \sigma _{0}\rangle = -2(N-1)\Gamma \left|\langle \sigma _{-}\rangle\right|^2 -  \Gamma\left[(2\bar{n}+1)\langle \sigma _{0}\rangle +1 \right],
\end{eqnarray}%
\begin{eqnarray}\label{s-}
\frac{d}{dt}\langle \sigma _{\pm}\rangle =\pm i\omega_{o} \langle \sigma _{\pm}\rangle
+\frac{(N-1)\Gamma}{2}\langle \sigma _{0}\rangle\langle \sigma _{\pm}\rangle
-  \frac{\Gamma(2\bar{n}+1)}{2}\langle \sigma _{\pm}\rangle.  
\end{eqnarray}%
The first term in the RHS of Eqs. (\ref{s0}), and the first and second terms in the RHS of Eq. (\ref{s-}) come from the evolution due the nonlinear Hamiltonian, while the last terms in both equations originate from the second term in the master equation (\ref{nlTerCla}), responsible for the dissipation mechanism.

As a first analysis, we will disregard dissipation and consider the evolution under the effective Hamiltonian (\ref{Hef2}) only. This consideration is valid for processes occurring in a time interval much shorter than the relaxation time, when the system becomes stationary. In this case, the dissipative terms in Eqs. (\ref{s0}) and (\ref{s-}) are disregarded, and the equations of motion for the mean values reduce to
\begin{equation}\label{s0b}
\frac{d}{dt}\langle \sigma _{0}\rangle = -2(N-1) \Gamma \langle \sigma _{+}\rangle\langle \sigma _{-}\rangle ,
\end{equation}%
\begin{equation}\label{s-b}
\frac{d}{dt}\langle \sigma _{\pm}\rangle =\pm i\omega_{o} \langle \sigma _{\pm}\rangle
+\frac{(N-1)\Gamma}{2}\langle \sigma _{0}\rangle\langle \sigma _{\pm}\rangle,
\end{equation}
that can be solved exactly, once recognizing that
\begin{equation}
R^2= 4\langle \sigma _{+}\rangle\langle \sigma _{-}\rangle+\langle \sigma _{0}\rangle^2 ,
\label{bloch}
\end{equation}
is a constant of motion. Thus, the equation of motion for $\langle \sigma _{0}\rangle$ becomes a first order separable differential equation,
\begin{equation}\label{s0c}
\frac{d}{dt}\langle \sigma _{0}\rangle = -\frac{(N-1)\Gamma}{2}  \left(R^2-\langle \sigma _{0}\rangle^2\right),
\end{equation}
whose solution is
\begin{equation}\label{solu}
\langle \sigma _{0}(t)\rangle=-R\tanh\left(\frac{(N-1)\Gamma}{2} R t - A\right) ,
\end{equation}
where $A=\tanh^{-1} \left(\langle \sigma _{0}(0)\rangle /R\right)$. Substituting Eq. (\ref{solu}) into Eq. (\ref{s-b}) and solving by integration we obtain the solution
\begin{eqnarray}\label{solu2}
\langle \sigma _{\pm}(t)\rangle=\langle \sigma _{\pm}(0)\rangle  e^{\pm i\omega_{o} t} \cosh (A) \, 
 \text{sech}  \left(\frac{(N-1)\Gamma}{2} R t- A\right ),
\end{eqnarray}
where $\cosh (A)=\left(1-\left(\langle \sigma _{0}(0)\rangle /R\right)^2\right)^{-1/2}$.

\baselineskip=11.2pt

\begin{figure}\label{f1}
  \center
  \includegraphics[scale = 1.2]{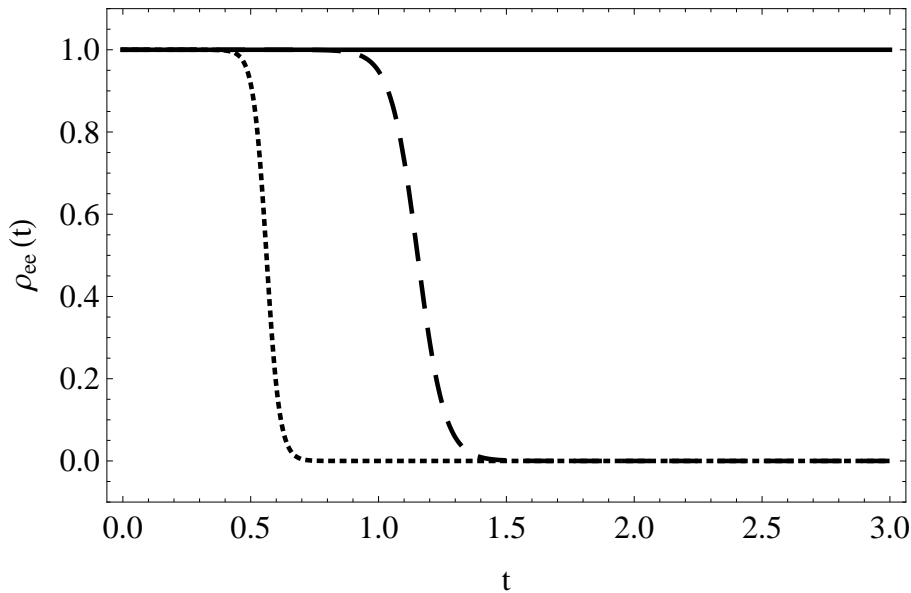}
  \caption{Excited state probability as function of time in the absence of dissipation. We set $\Gamma = 1$. Solid line: $N=1$; dashed line: $N=20$; dotted
line: $N=40$.}
\end{figure}

In order to illustrate the theory, the excited state probability  $\rho_{ee}=(1+\left\langle \sigma_{0}\right\rangle)/2$ is plotted in Fig. \ref{f1} as function of time for $\rho_{ee} (0)=1$, $\Gamma =1$, and for several values of $N$. Note that by having disregarded the dissipative terms there is no spontaneous decay. So, in the situation at hand the generic particle does not  decay spontaneously when it is initially set in the excited state, namely $\rho_{ee}(0)=1$, which is an  unstable  equilibrium  point, since it lies in the excited state with zero dipole moment. The curves in Fig. \ref{f1} are obtained for $\rho_{ee}(0)$ slightly smaller than $1$. The decay profile is due the mean field created by the other particles, and not by dissipative Sudarshan-Lindblad terms. We note that the mean energy of the system $\langle \hat{E}\rangle=N\hbar \omega_{o} \langle \sigma _{0}(t)\rangle$ is not a conserved quantity, it depends on the initial state, $\rho_{ee}(0)$, being less than 1, when the spins tend to align along the effective polarization mean field.

In fact, the equations of motion generated by the Hamiltonian (\ref{Hef2}) are equivalent to those describing the superradiant emission by a collection of two-level atoms in the framework of a semi-classical radiation theory \cite{stroud}, described by a single-particle mean-field Hamiltonian for the individual emitters as pointed out in Ref. \cite{salomon1}.

\baselineskip=12pt

\begin{figure} \label{f2}
  \center
  \includegraphics[scale = 1.2]{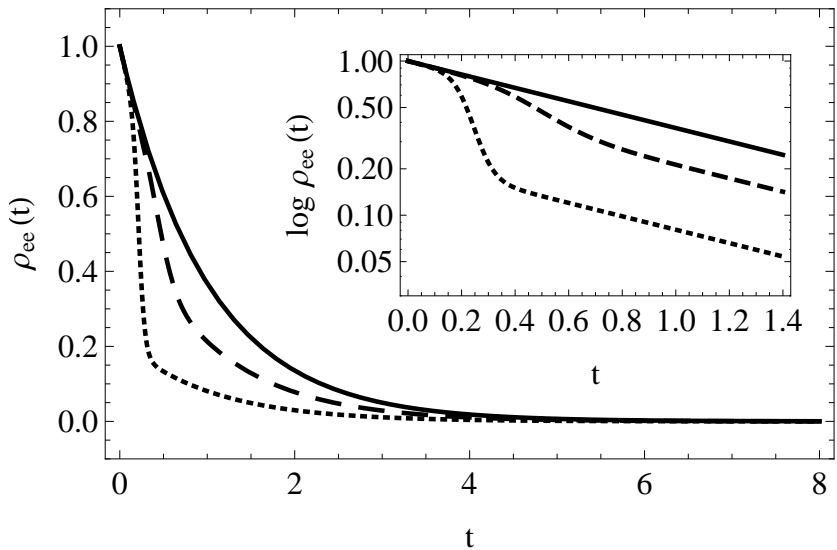}
  \caption{Excited state probability as function of time including dissipation. We set $\Gamma =1$ and $\bar{n} =0$. Solid line: $N=1$; dashed line: $N=20$; dotted line: $N=40$.}
\end{figure}

\begin{figure} \label{f3}
  \center
  \includegraphics[scale = 1.2]{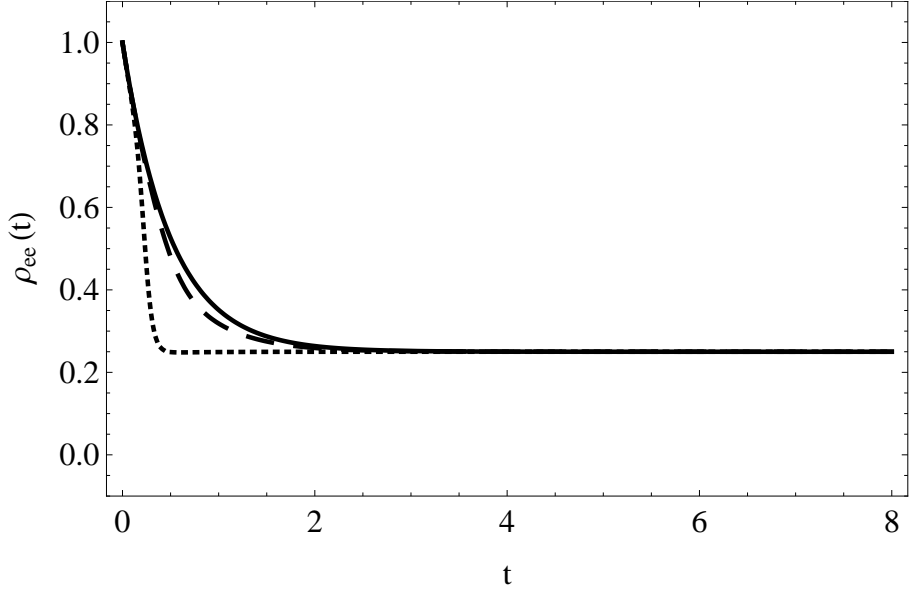}
  \caption{Excited state probability as function of time including dissipation. We set $\Gamma =1$ and $\bar{n} =0.5$. Solid line: $N=1$; Dashed line: $N=20$; Dotted line: $N=40$.}

\end{figure}

We turn now to Eqs. (\ref{s0}) and (\ref{s-}). For $N=1$ the system reduces to
a decoupled set of linear equations whose solutions are
\begin{equation}
\left\langle \sigma _{0}(t)\right\rangle=\left\langle \sigma _{0}(0)\right\rangle
e^{ -(2\bar{n}+1)\Gamma \,t}+\frac{e^{ -(2\bar{n}+1)\Gamma \,t}-1}{2\bar{n}+1},
\end{equation}
\begin{equation}
\left\langle \sigma _{\pm}(t)\right\rangle=\left\langle \sigma _{\pm}(0)\right\rangle
e^{\pm i\omega_{o}t}e^{-\frac{1}{2}(2\bar{n}+1)\Gamma \,t},
\end{equation}
that contain a transient exponential decay and at long times attain the stationary values
$\left\langle \sigma _{0}(\infty)\right\rangle= -1/(2\bar{n}+1)$ and
$\left\langle \sigma _{\pm}(\infty)\right\rangle=0$, respectively.
For the system of Eqs. (\ref{s0})-(\ref{s-}), when $N>1$, the analytical solution is possible only for the
stationary state, otherwise a numerical calculation is necessary. The stationary solution
is obtained by setting $d\langle \sigma _{0}\rangle /dt = d\langle \sigma _{\pm}\rangle /dt=0$,
and solving the resulting algebraic equations whose solutions are identical to the case $N=1$.
Therefore, the stationary state does not depend on the number of particles and is determined solely by the reservoir temperature through the parameter $\bar{n}$ that depends on the absolute temperature $T$. The deviation from an exponential decay is due to the nonlinear terms and it occurs at transient times. To illustrate the interplay between nonlinear and dissipative effects, in Fig. \ref{f2} we plotted $\rho_{ee}$ as function of time
for $\rho_{ee}(0)=1$, $\Gamma =1$, and setting the reservoir temperature at $0$K ($\bar{n}=0$),
for several values of $N$. For $N=1$ we have a purely exponential decay (solid line) while for $N>1$ the deviation from
the exponential decay at transient times due the nonlinear mean-field is remarkable (dashed and doted lines). This deviation is evidenced in the inset within the Fig.~\ref{f2} drawn in logarithmic scale. Figure \ref{f3} is the same as Fig. \ref{f2} but for $\bar{n}=0.5$; we note that at higher temperature the decay is faster and $\rho_{ee} (t)$ attains an asymptotic nonzero value for the stationary state that will depend on the temperature on
the value of $\bar{n}$.
\vspace{2mm}

\section{Beyond the mean-field approximation}{\label{section5}}
%
Here we present a brief discussion on how to go beyond the single-particle mean-field approximation, which consists in the factorization $\rho_{2}(1,2)=\rho_{1}(1)\rho_{1}(2)$, that disregards the two-particle correlations that exist, either in the initial conditions, or
being generated by the interaction with the environment, if the gas is quite diluted, as the system of particles that evolve dynamically in time. Under stricter conditions, the correlations
between the particles must be considered in order to describe the dynamics more faithfully, and this means that we have to add, at least, a two-particle operator $\chi (1,2)$, such that
\be
\rho_{2}(1,2)=\rho_{1}(1)\rho_{1}(2) + \hat{\chi} (1,2),
\label{chi}
\ee
having the properties ${Tr}_2 \hat{\chi} (1,2) = {Tr}_1 \hat{\chi} (1,2) = 0$, where $Tr_{k}$ means calculating the trace over the degrees of freedom of particle $k$ . The operator
$\hat{\chi} (1,2)$ contains all kind of two-particle correlations, it
can be introduced within the initial conditions for $\rho_{2}(1,2)$, and its evolution in time will be ruled by the master equation. However, according to the BBGKY equation
(\ref{masterKterClaF}), if we set in its left hand side (LHS) the time derivative of a two-particle density operator, the RHS will contain a dependence on a three-particle density operator, $\rho_{3}(1,2,3)$, and in order to solve the differential equation one should go a step further, namely set the time derivative on the LHS for a three-body operator and we shall get on the RHS of the equation a dependence on a four-particle operator. This procedure goes on until we set the exact equation for the $N$-particle problem. As this procedure increases in complexity, we have adopted the quite simple single-particle mean-field approximation. To go beyond this approximation, and still to extend the BBGKY equation that involves two-particle correlations, we can adopt the following ansatz for the three-particle density operator,
\begin{eqnarray}\label{rho3}
\rho_{3}(1,2,3) \approx \rho_{1}(1) \rho_{2}(2,3) + \rho_{1}(2) \rho_{2}(1,3) + \rho_{1}(3) \rho_{2}(1,2)
- 2 \rho_{1}(1) \rho_{1}(2) \rho_{1}(3),
\end{eqnarray}
which satisfies the requirements $Tr_{3} \rho_{3}(1,2,3) = \rho_{2}(1,2)$,
$Tr_{2} \rho_{3}(1,2,3) = \rho_{2}(1,3)$ and $Tr_{1} \rho_{3}(1,2,3) = \rho_{2}(2,3)$.

By introducing the correlation operator (\ref{chi}) in Eq.~(\ref{rho3}) we get
\begin{eqnarray}\label{rho2}
\rho_{3}(1,2,3) \approx \rho_{1}(1) \rho_{1}(2) \rho_{1}(3) + \rho_{1}(1) \hat{\chi} (2,3) + \rho_{1}(2) \hat{\chi} (1,3) + \rho_{1}(3) \hat{\chi} (1,2).
\end{eqnarray}
Thus, the BBGKY equation reduces to a closed system of coupled differential equations involving the
one-body density operator and the two-body correlation operator. The set of the dynamical equations for $\rho_{1}(k)$
and $\hat{\chi} (j,k) $ is still not easy to solve analytically, one could adopt a further
simplification, that is known as the linear response approximation, where one looks for the correlations
nearby a stationary or equilibrium state of the system, represented by the one-particle operator obtained under the mean-field approximation, namely, $\rho_{1}^{st}(k)$. This ansatz reduces the density operator (\ref{rho2}) to
\begin{eqnarray}\label{rhos}
\rho_{3} (1,2,3) \approx \rho_{1}^{st}(1) \rho_{1}^{st}(2) \rho_{1}^{st}(3) + \rho_{1}^{st}(1) \hat{\chi}(2,3) + 
\rho_{1}^{st}(2) \hat{\chi} (1,3) + \rho_{1}^{st}(3) \hat{\chi} (1,2),
\end{eqnarray}
and the BBGKY becomes a dynamical equation for the correlation operator $\hat{\chi} (j,k)$ only.

In addition, the statistical operator for one and two particles can be written in terms of the Pauli
matrices $\mbox{\boldmath$\sigma$} =(\sigma_{x},\sigma_{y},\sigma_{z})$ and the unit operator $I$, which form a basis for
$2\times 2$ matrices \cite{blum}. Thus, for the one-particle density operator we have
\begin{equation}
\rho_{1}=\frac{1}{2} \left[ I + \left\langle \mbox{\boldmath$\sigma$}\right\rangle \cdot \mbox{\boldmath$\sigma$}  \right],
\end{equation}
where $\langle \mbox{\boldmath$\sigma$}\rangle =(\langle \sigma_{x}\rangle,\langle \sigma_{y}\rangle,\langle \sigma_{z}\rangle)$ is the so-called Bloch vector \cite{Breu02,milb,blum}; while the two-particle density operator, expanded in a basis formed by $16$ two-particle spin operators, $\left\{
I_{1}\otimes I_{2},I_{1}\otimes \sigma_{i}(2), \sigma_{i}(1)\otimes I_{2},  \sigma_{i}(1)\otimes \sigma_{j}(2)  \right\}$, has the following form \cite{blum}
\begin{eqnarray}\label{dois}
\rho_{2}(1,2)=\frac{1}{2^{2}}\bigg[ I_{1}\otimes I_{2}+ \left\langle \mbox{\boldmath$\sigma$}(1)\right\rangle \cdot\mbox{\boldmath$\sigma$}(1)
\otimes I_{2} + 
  I_{1}\otimes \left\langle \mbox{\boldmath$\sigma$}(2)\right\rangle\cdot \mbox{\boldmath$\sigma$}(2) + \sum_{ij} B_{ij} \sigma_{i} (1)\otimes
\sigma_{j}(2) \bigg],
\end{eqnarray}
with $B_{ij}=\left\langle \sigma_{i}(1)\sigma_{j}(2)\right\rangle $ being the two-particle correlations coefficients. By writing
\begin{equation}\label{bij}
B_{ij}=\left\langle \sigma_{i} (1)\right\rangle \left\langle \sigma_{j}(2)\right\rangle + 4 \chi_{ij},
\end{equation}
the elements $\chi_{ij}$ are identified with the entries of the matrix of covariances. The  factor $4$ was included for convenience.  Inserting the Eq. (\ref{bij}) into Eq. (\ref{dois}), the Eqs. (\ref{chi}) and (\ref{rho2}) are recovered, with the identification $\hat{\chi}_{2}(1,2)=\chi_{ij} \sigma_{i} (1)\otimes \sigma_{j}(2)$.
%
\section{Summary and conclusions}{\label{section6}}
%
The aim of this paper is to be pedagogical for advanced undergraduate and graduate students by introducing some techniques and approximations that arise in the physics that involves dissipation and nonlinearity in many-particle systems. Here we have considered a system $\mathcal{S}$ of $N$ independent spin-1/2 particles interacting with a thermal
reservoir $\mathcal{R}$. We started with the dynamical quantum Sudarshan-Lindblad master equation for the $N$-particle density operator. From this point on we derived the quantum version of the BBGKY
hierarchy of $K$ equations for $1 \leq K < N$. Then, by breaking the hierarchy we neglected
interparticle correlations, obtaining thus an equation for a representative single-particle.
This is equivalent to introduce the mean-field
approximation in the  equation for an $N$ particle system. This scenario
leads to a quite similar Sudarshan-Lindblad equation for the single-particle density operator; however, a main difference
arises: the Hamiltonian is now nonlinear, although the Sudarshan-Lindblad superoperator,
responsible for the dissipation and decoherence in $\mathcal{S}$, remains unaffected.
In order to provide a useful tool to students interested in many-particle problems,
we have discussed the physical approximations involved, we derived the analytical solution
for the nonlinear mean-values operators, whose motion is generated by the nonlinear Hamiltonian without dissipative terms, and illustrate numerically the case with dissipation. We also included a discussion on how to go beyond the mean-field approximation by introducing two-particle correlations.

Finally, we believe that this paper can be useful for the reader interested in doing research in theoretical physics of many-particle systems. For some related papers on nonlinearity and dissipation in many-particle systems, see Refs. \cite{x1,x2,x3,x4,x5,x6,x7,x8,x9}.

\bigskip
%
%



\newpage

\bigskip

\end{document}